Review

# User-Centered Design for Digital Patient-Navigation Tools in Oncology: Scoping Review

Saba Kheirinejad[1], PhD; Brianna M White[1], DrPH, MPH; Parnian Kheirkhah Rahimabad[1], MD, PhD; Janet A Zink[1], PhD, MS; Soheil Hashtarkhani[1], PhD; Fekede Asefa Kumsa[1], PhD; Rezaur Rashid[1], PhD; Lokesh Chinthala[1], PhD; Christopher L Brett[2], MD; Robert L Davis[1], MD, MPH; David L Schwartz[3,4], MD; Arash Shaban-Nejad[1], PhD, MPH

[1]Center for Biomedical Informatics, Department of Pediatrics, College of Medicine, University of Tennessee Health Science Center, Memphis, TN, United States

[2]Graduate School of Medicine, University of Tennessee at Knoxville, Knoxville, TN, United States

[3]Departments of Radiation Oncology, College of Medicine, University of Tennessee Health Science Center, Memphis, TN, United States

[4]Departments of Preventive Medicine, College of Medicine, University of Tennessee Health Science Center, Memphis, TN, United States

**Corresponding Author:**
Arash Shaban-Nejad, PhD, MPH
Center for Biomedical Informatics, Department of Pediatrics, College of Medicine
University of Tennessee Health Science Center
50 N Dunlap St
Memphis, TN, 38103
United States
Phone: 1 9012875863
Email: ashabann@uthsc.edu

## Abstract

**Background:** Navigation programs for patients with cancer improve access and continuity of care, yet their digital transformation is often limited by poor usability and inadequate uptake. Applying user-centered and human-centered design (UCD/HCD) principles may close this gap, but the extent to which such design methods are used and evaluated in oncology navigation tools remains unclear.

**Objective:** This scoping review identifies how UCD/HCD principles have been, and should be, applied in developing and implementing digital health tools for navigation for patients with cancer.

**Methods:** A scoping review was conducted following PRISMA-ScR (Preferred Reporting Items for Systematic Reviews and Meta-Analyses extension for Scoping Reviews) and Joanna Briggs Institute guidance. A total of 7 databases (PubMed/MEDLINE, Scopus, IEEE Xplore, Web of Science, Embase, ACM Digital Library, and CINAHL) were searched for English-language articles published between January 2015 and July 2025. Eligible studies reported original, peer-reviewed research on digital or mobile health interventions linked to cancer navigation and documented at least 1 UCD/HCD activity. Two reviewers independently screened records and charted data on context, target users, functions, tool modality, design phase, methods, and outcomes. Findings were synthesized descriptively and thematically.

**Results:** A total of 36 studies met the inclusion criteria. Findings were organized into 4 domains: study characteristics, navigation functions and digital modalities, design processes and methods, and UCD/HCD application. Iterative prototyping and usability testing were the most common, while participatory design and implementation evaluation were underused.

**Conclusions:** UCD/HCD approaches enhance usability and patient relevance of digital cancer navigation tools. However, their application remains limited across cancer types, regions, and functions. Broader stakeholder participation and evaluation beyond usability are needed to strengthen coordination, equity, and sustainability in cancer care.

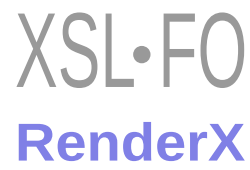

## Introduction

Cancer remains a leading health challenge, with an estimated 35 million new cases to be diagnosed globally by 2050 [1]. Given ever-aging and evolving populations, the intricate nature of cancer care can pose significant challenges to patients and those caring for them [2]. Patient navigation strategies have emerged to eliminate barriers and improve access to care [3]. Community-based patient navigation programs provide personalized support to patients, families, and caregivers across the cancer care continuum [3,4].

Navigation programs improve care engagement, enhance adherence and treatment completion, reduce diagnostic delays, support survivorship care, and improve overall patient quality of life and satisfaction. Moreover, these programs can help patients overcome language barriers, schedule care appointments, and respond to social needs (ie, lack of transportation, childcare, etc). Patient navigation programs also empower patients by helping them better manage their own medical information and promote active participation in their care, including the power to make informed decisions [4,5].

Despite these benefits, patient navigation program design and delivery vary widely and lack standardization. This complicates multidisciplinary care coordination and introduces additional challenges to already complex health care systems, including financial sustainability concerns that may present an unmanageable burden for many care centers [5]. Addressing these issues requires a comprehensive understanding of the sociotechnical aspects of current technologies and workflows. To optimize the reach and effectiveness of patient navigation programs, there is growing interest in leveraging digital health technologies, including mobile health (mHealth) solutions.

Digital health broadly refers to the use of information and communication technologies to support health care delivery, including web-based platforms, mobile apps, telehealth systems, remote monitoring tools, and data-driven decision support technologies [6]. In oncology, digital health tools supporting patient navigation span multiple modalities, such as web portals, mobile apps, hybrid systems, and, increasingly, artificial intelligence–enabled or conversational tools [7]. Defined by the World Health Organization as “medical and public health practice supported by mobile devices,” mHealth encompasses a wide range of tools, including mobile apps, wearables, sensors, and communication platforms designed to improve health outcomes and promote patient-centered care [8]. In oncology settings, mHealth interventions have been used to deliver real-time patient alerts, streamline manual administrative tasks, and improve communication between patients and care teams [9]. Accordingly, this review adopts a broad digital health lens, situating mHealth within a wider ecosystem of technologies used to support navigation for patients with cancer.

Qualitative studies have found that oncology navigators particularly value digital features that reduce administrative workload and facilitate timely intervention (ie, automatic appointment reminder systems, real-time patient messaging, etc). The findings emphasize the importance of care coordination for patients with cancer undergoing chemotherapy and the critical role of technology in improving this process by enhancing communication and access to information for patients [9]. However, many digital tools fail to simultaneously meet the needs of patients and care teams due to poor design, low usability, and a lack of user engagement. Patients often disengage from apps that are not intuitive, while care navigators may find systems incompatible with clinical workflows. To address these issues, there is increasing recognition of the importance of user-centered design (UCD) and human-centered design (HCD).

UCD is a design philosophy that focuses on placing the end user at the center of the design process [10]. In some contexts, the terms HCD and UCD are used interchangeably. However, HCD is often used to emphasize a broader consideration of impacts on various stakeholders, beyond just the direct users. A UCD/HCD approach, which focuses on understanding user needs, engaging stakeholders, and creating tailored solutions [11], can help address these complexities and improve care delivery [12]. These approaches prioritize the needs, preferences, and lived experiences of end users (ie, patients, caregivers, and providers) through iterative, participatory design processes that foster usability, engagement, and real-world applicability [13,14].

Specifically, cancer care is a complex and emotionally demanding process that involves not only patients but also caregivers and clinicians. Patients often face multiple treatments, fluctuating symptoms, and interactions with numerous health care professionals across different systems. Caregivers experience emotional stress and logistical burdens, while clinicians manage heavy workloads and complex decision-making demands. These challenges can lead to physical strain, emotional distress, cognitive fatigue, and variable energy levels, all of which influence how each group engages with digital tools. Therefore, designing tools that are intuitive, accessible, and emotionally supportive is essential for all stakeholders, as this can enhance usability, encourage sustained engagement, and foster coordinated cancer care.

Given the potential for UCD and HCD to improve their effectiveness, there is a need to systematically map the literature on how these design principles have been applied in this context. This scoping review aims to identify and characterize how UCD and HCD approaches have been applied across the development and implementation of digital health and mHealth tools to support navigation for patients with cancer. Rather than assessing effectiveness, the review examines how UCD/HCD has been operationalized across four interrelated dimensions: (1) study characteristics, (2) navigational functions and digital modalities, (3) design processes and methods, and (4) UCD/HCD application. These dimensions were selected because design choices influence not only usability, but also stakeholder inclusion, workflow integration, scalability, and equity. By evaluating current practices against these dimensions, this review seeks to clarify where UCD/HCD efforts have been concentrated, where gaps remain, and how future digital navigation tools might better align with the complex realities of cancer care.

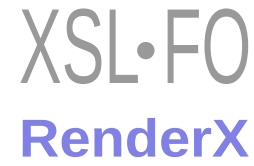

## Methods

### Overview

This review was guided by the Joanna Briggs Institute methodology [15] for scoping reviews and the PRISMA-ScR (Preferred Reporting Items for Systematic Reviews and Meta-Analyses extension for Scoping Reviews) guidelines (Multimedia Appendix 1) [16]. Ethical approval was not applicable to this study, as it involved secondary analysis of previously published, peer-reviewed literature and did not include human participants.

### Search Strategy

A comprehensive literature search was undertaken to map how UCD and HCD principles have been used when developing or implementing digital and mHealth tools for navigation for patients with cancer. The search strategy was developed iteratively by the study team; no formal librarian consultation was conducted. The review incorporated peer-reviewed studies that were identified from 7 electronic databases (PubMed/MEDLINE, Scopus, IEEE Xplore, Web of Science Core Collection, Embase [Ovid], ACM Digital Library, and CINAHL [EBSCO]) and were published between January 1, 2015, and July 30, 2025. The search was conducted on July 30, 2025, and the strategy combined the following three concept blocks joined by the Boolean operator AND: (1) cancer navigation terminology ("cancer navigation," "patient navigation," "care coordination," "care transitions," "cancer care pathway," "cancer journey support"), (2) digital and mHealth terminology ("digital health," "mobile health," "mHealth," "eHealth," "mobile application," "wearable," "remote monitoring," "telehealth," "teleoncology," "health technology"), and (3) UCD/HCD terminology ("user-centered design," "human-centered design," "UCD," "HCD," "participatory design," "person-centered care" combined with design, "usability testing"). Database-specific controlled vocabulary (eg, Medical Subject Headings and Emtree) was incorporated where available, and search strings were adapted to the syntax of each platform.

Full, database-specific search strategies and dates of execution are provided in Multimedia Appendix 2.

### Eligibility Criteria

Studies were included for review if they met all of the following criteria: (1) published between January 1, 2015, and July 30, 2025, (2) described the development or implementation of 1 or more cancer care navigation-related functions, including care coordination, access facilitation, communication between patients and care teams, appointment management, psychosocial support, financial or logistical assistance, or guidance across the cancer care continuum, (3) reported at least 1 UCD or HCD component (ie, needs assessment, co-design workshop, iterative prototyping, and formal usability testing), (4) involved patients with cancer of any age, their caregivers, or navigation personnel, (5) were original, peer-reviewed research articles or design case studies, and (6) were available in the English language. Studies were excluded if they evaluated oncology digital tools unrelated to navigation, lacked any description of UCD/HCD processes, examined nononcology populations, or appeared as conference abstracts, dissertations, protocols, commentaries, or other gray literature.

### Study Selection

All search results were imported into Covidence [17] for screening. Duplicates were removed automatically and confirmed manually. The screening process was conducted in two sequential stages: (1) a primary title and abstract screening followed by full-text review of potentially eligible studies and (2) a secondary bidirectional snowballing process, which involved recursively pursuing relevant references cited in the retrieved literature (backward snowballing) as well as identifying studies that cited the included articles (forward snowballing).

Two reviewers (SK and BMW) independently screened titles and abstracts to assess relevance to the eligibility criteria; disagreements were resolved through discussion, with a third reviewer available for arbitration if needed. Studies meeting the inclusion criteria were advanced to the full-text review stage, where the same dual-review procedure was applied.

### Definitions and Conceptual Framework

In this review, digital health refers broadly to the use of information and communication technologies to support health care delivery, including web-based platforms, mobile apps, telehealth systems, remote monitoring tools, and data-driven decision support technologies. mHealth represents a subset of digital health technologies supported by mobile devices.

UCD and HCD are design approaches that emphasize the active involvement of end users and stakeholders throughout the design process to ensure that technologies align with user needs, values, and contexts. In this review, UCD/HCD applications were examined across multiple design phases, including discovery, design, prototyping, evaluation, and implementation.

Design methods refer to specific activities or techniques used to engage users or stakeholders, such as interviews, focus groups, co-design workshops, or usability testing. These design methods may be applied at different design phases and are not mutually exclusive.

Participatory methods are defined as design approaches that involve users or stakeholders as active contributors to the design process rather than solely as subjects of evaluation.

Implementation, as used in this review, refers to the deployment and integration of digital navigation tools within real-world care settings, including considerations related to feasibility, workflow integration, training, and sustainability.

### Data Extraction

Data were extracted using a pilot charting form in Microsoft Excel. Two reviewers (SK and BMW) independently recorded bibliographic details (author, year, country, and journal), study design and methods, participant characteristics (cancer type, age range, and navigator role), digital tool attributes (platform, functionality, and stage of deployment), UCD/HCD activities undertaken, navigation tasks supported, and reported outcomes (usability measures, patient engagement, navigator workload,

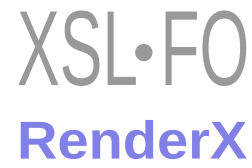

and clinical indicators). Discrepancies were reconciled by consensus.

### Data Analysis

Thematic analysis was conducted through a qualitative synthesis of the studies presented in Table S1 in Multimedia Appendix 2. Each article was reviewed in full to identify key concepts, use cases, and reported outcomes relating to the application of UCD/HCD in digital cancer navigation. An inductive approach was used to identify patterns emerging directly from the data, which were then organized into the following four overarching domains reflecting the clinical and methodological dimensions of UCD/HCD: (1) study characteristics, (2) navigation functions and digital modalities, (3) design processes and methods, and (4) conceptualizations of cancer navigation through UCD/HCD.

Domains were refined iteratively until they provided a coherent, nonoverlapping structure for grouping the findings. Alongside the qualitative synthesis, quantitative frequencies (publication year, geography, cancer type, and digital modality) were summarized in descriptive tables and figures to map the landscape of evidence.

## Results

### Overview

A total of 36 studies met the inclusion criteria for review, as shown in Figure 1. Of the 192 records screened for inclusion through the primary title and abstract review, 176 records were excluded for failure to meet the eligibility criteria. In total, 16 studies met the inclusion criteria and were retained for qualitative synthesis. Bidirectional snowballing of these records resulted in the inclusion of 20 additional articles.

**Figure 1.** PRISMA-ScR flow diagram of literature search and screening.

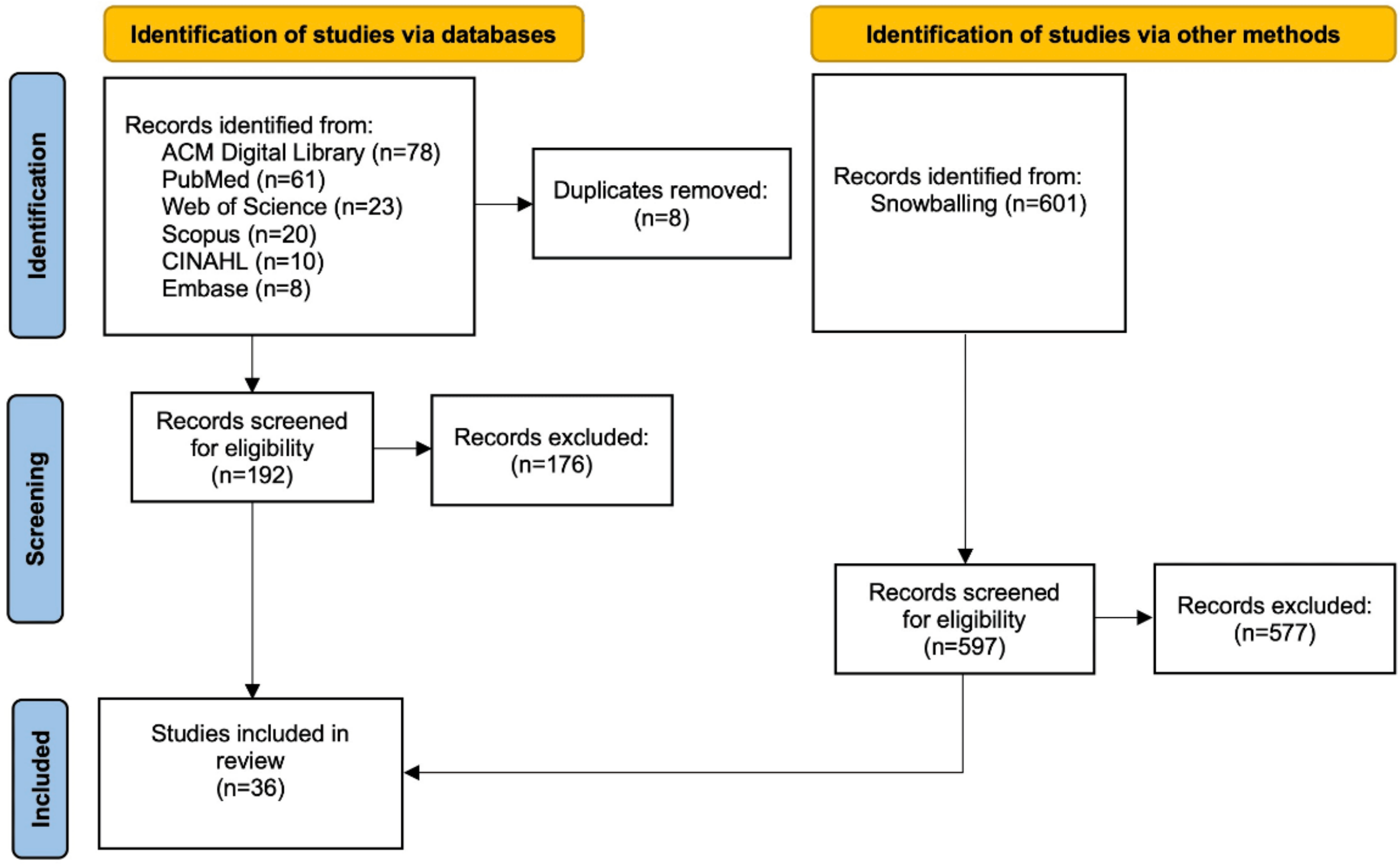

Study characteristics are summarized in Table S1 in Multimedia Appendix 2. Below, findings are organized around 4 domains that emerged from qualitative synthesis of the extraction matrix.

Table 1 presents a high-level synthesis of the thematic findings from the included studies, summarizing reported UCD/HCD design processes, functional scope, and digital modalities, and evaluation approaches and impacts in oncology patient navigation tools. The table highlights commonly used methods, dominant functional focuses, and key gaps identified across the literature. Descriptive study characteristics, such as cancer types, geographic distribution, and study settings, are provided separately in the text and Table S1 in Multimedia Appendix 2 to contextualize the evidence base.

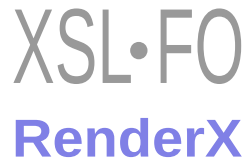

**Table 1.** Summary of key findings on user-centered design (UCD)/human-centered design (HCD) for digital patient navigation in oncology.

| Domain | Key themes | Key findings |
|---|---|---|
| Design processes (UCD/HCD phases, design methods, and stakeholder engagement) | • Iterative prototyping and usability testing were most common.<br>• Discovery and implementation phases rarely included.<br>• Interviews and surveys were most frequently used, few participatory or co-design workshops.<br>• Engagement was largely patient-only; minimal involvement of clinicians, navigators, or caregivers. | • UCD applied primarily to midcycle for usability evaluation.<br>• Participatory methods underused, limiting innovation and sustainability. |
| Functional scope of tools (navigation purpose and digital modality) | • Web portals and mobile apps were overwhelmingly used; focused primarily on education and communication.<br>• Limited coverage of financial, logistical, or psychosocial barriers.<br>• Few integrated or multimodal systems. | • Tools improved patient knowledge and engagement.<br>• Narrow functional range; informational focus was most common.<br>• Opportunities to integrate financial, transportation, and emotional-support features remain. |
| Evaluation and impact (navigation outcomes and author reflections) | • Usability and satisfaction were the most used metrics.<br>• Few studies assessed navigation effectiveness, coordination, or clinical outcomes.<br>• Equity, timeliness, and continuity rarely evaluated.<br>• Some studies conceptually broadened navigation toward empowerment and communication. | • Tools perceived as usable but with limited real-world evidence of improved care continuity.<br>• Shift needed from acceptability to system-level effectiveness.<br>• UCD reframes navigation as multidimensional (communication, coordination, and empowerment). |

## Study Characteristics

Across the 36 included studies, most focused on all adult cancers (n=16, 44.4%) or specifically on breast cancer (n=8, 22.2%). Fewer addressed pediatric populations (n=3, 8.3%), lung (n=3, 8.3%), or prostate cancer (n=2, 5.6%), while single-study contexts included brain, cervical, chronic myeloid leukemia, colorectal, esophageal, head and neck, melanoma and renal cell carcinoma, and multiple myeloma (n=1, 2.8% each). This distribution reflects a strong emphasis in the current literature on adult and breast cancer contexts, with limited representation of other cancer types and life stages.

Regarding target user groups, most digital tools were designed for patients (33/36, 91.7%) and clinicians or providers (25/36, 69.4%), while caregivers (6/36, 16.7%) and patient navigators (2/36, 5.6%) were considered in just 2 studies. This distribution highlights a strong emphasis on patient-facing tools in the current literature, with comparatively limited attention to caregivers and patient navigators. For example, OncoKompas [13], an app codeveloped with cancer survivors, enables users to monitor their quality of life through patient-reported outcomes and receive automated, personalized feedback and supportive care recommendations, illustrating how tools are often designed directly with patients in mind to provide personalized self-management support.

Geographically, nearly half of the studies were conducted in the United States (17/36, 47.2%), followed by the Netherlands (7/36, 19.4%) and Canada (3/36, 8.3%). Fewer studies were conducted in Africa, Germany, and Sweden (2/36, 5.6% each), and from Australia, Iran, Norway, and Taiwan (1/36, 2.8% each). The predominance of studies from North American and European settings underscores a need for more geographically diverse applications of UCD/HCD approaches to enhance their generalizability and global relevance in cancer navigation.

## Navigation Functions and Digital Modalities

### *Navigation Purpose*

According to our findings, navigation functions were most frequently conceptualized as education and information provision (23/36, 63.9%). Closely following were overall care system navigation or access to care and patient-to-provider communication (21/36, 58.3% each), along with care coordination (20/36, 55.6%). Less frequently, studies reported functions related to emotional or psychosocial support (11/36, 30.6%) and appointment scheduling and tracking (7/36, 19.4%). Additionally, functions included patient-to-patient communication (3/36, 8.3%), symptom management (2/36, 5.6%), and financial or insurance navigation, medication support, and transportation or logistics (1/36, 2.8% each).

These results show a predominant focus on education, system navigation, and communication functions, with limited attention to supportive services, financial or logistical navigation, and peer interaction. For example, Mendu et al [18] created and tested a culturally tailored virtual patient educator for Hispanic women to improve cervical cancer and human papillomavirus education using an iterative, UCD. The *CMyLife* platform presented by Ector et al [14] was codeveloped to help patients interpret their own test results and take action across their care journey, reframing navigation as patient empowerment rather than simple wayfinding. Similarly, digital personal health network systems supported chemotherapy patients by enhancing communication with providers and improving care coordination, while usability studies identified needed refinements in data entry and connectivity [9,19]. Moreover, interventions such as *iNNOVBC* presented by Mendes-Santos et al [20] highlighted the underexplored potential of navigation tools to provide psychological support, positioning digital navigation to address care gaps in survivorship.

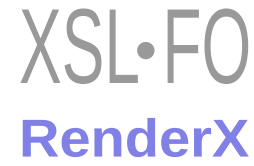

### *Type of Digital Tools*

Web platforms were the most common type of proposed digital tools, reported in 18 (50%) studies, closely followed by mobile apps (17/36, 47.2%). A smaller proportion presented only a conceptual design without testing a specific tool (3/36, 8.3%), while software prototypes and telehealth or video tools were each reported in a single study (1/36, 2.8%).

This distribution suggests that research has largely concentrated on familiar digital formats (ie, web platforms and mobile apps), with comparatively limited integration of newer modalities such as conversational agents or artificial intelligence–driven chatbots, which were categorized as digital health modalities within this review. For example, the *MijnAVL* [21] platform was iteratively developed with survivor input as a patient-facing web portal to enhance usability and provide accessible educational content. Similarly, the *StressProffen* [22] platform illustrates the use of mobile apps to deliver psychosocial support. Broader approaches also emerged, such as a human-centered mobile and web application (*mPCL*) [23] in Tanzania, which extended functionality beyond patient-facing use to support coordinated palliative care among patients, caregivers, and local health workers.

## UCD/HCD Phase and Methods Used

Across studies included, UCD/HCD approaches were reported at multiple phases of the design cycle, including discovery, design and prototyping, testing, and implementation. Rather than following a linear or standardized sequence, most studies reported applying UCD/HCD selectively at 1 or more phases, with considerable variability in timing and depth.

In this review, discovery refers to early-stage activities aimed at understanding user needs, care contexts, and system requirements through methods such as interviews, focus groups, and needs assessments. Design and prototyping encompass the iterative creation of low- to high-fidelity solutions, including wireframes, mock-ups, and functional prototypes, often developed through co-design with stakeholders. Testing includes usability evaluations and pilot studies conducted to assess feasibility, acceptability, and workflow fit in controlled or limited real-world settings. Implementation indicates the deployment and integration of a digital navigation tool into routine clinical workflows or operational care settings, where real users engage with the system as part of care delivery. Distinguishing these phases allowed us to examine the maturity of UCD/HCD application across studies and identify gaps in extending user-centered practices into real-world deployment.

Discovery-phase activities, such as needs assessments, contextual inquiry, and early exploration of user workflows, were reported in fewer than half of the studies (17/36, 47.2%). These approaches were typically used to identify barriers to care coordination, information gaps, or workflow challenges faced by patients, caregivers, or clinicians. As examples in early discovery, Carr et al [12] examined barriers in the workflows of patient navigators supporting patients with breast cancer and explored how an mHealth app could help improve their care delivery and coordination, Ankrah et al [24] interviewed childhood cancer survivors and caregivers to explore survivorship experiences and offered sociotechnical design insights to better support young survivors, and Lau and Loiselle [25] examined oncology nurses' views of an evidence-based eHealth tool, highlighting that incorporating their perspectives makes eHealth tools valuable for patient care and nursing practice.

Design and prototyping phases were more frequently represented (19/36, 52.7%), often involving the development of functional prototypes and working systems. Studies such as *CMyLife* [14] and *iNNOVBC* [20] illustrate how patient-centered platforms were developed and iteratively refined through user feedback, leading to improvements in navigation, personalization, and care coordination features. These examples highlight the use of UCD/HCD primarily to refine near-functional systems and support real-world implementation.

Evaluation and testing activities were the most consistently reported component of UCD/HCD application. Testing was used in 63.8% (23/36) of studies, reflecting a strong emphasis on evaluating and refining tools once developed. As an example in usability testing, the *ASyMS* platform developed by Moradian et al [26], a remote monitoring system, underwent think-aloud usability testing with patients to refine symptom reporting functions before moving toward broader clinical use.

Implementation was the least represented phase (7/36, 19.4%), indicating that relatively few studies extended user-centered approaches into deployment, integration, or evaluation of digital navigation tools within routine clinical workflows. Studies such as Mittal et al [27] and Nabelsi and Plouffe [28] demonstrated that user feedback during implementation revealed system-level challenges, including training needs and resource constraints. These findings underscore the importance of extending UCD/HCD beyond development and evaluation into real-world deployment contexts.

Overall, UCD/HCD approaches were mostly used in prototyping and evaluation phases, with less frequent attention to early discovery and downstream implementation in cancer navigation. Their potential for shaping problem definition and ensuring sustainable implementation remains less frequently explored in available literature.

A variety of UCD/HCD design methods were reported across the included studies, with 80% (29/36) of the studies reporting the use of more than 1 method across the design process. Interviews were the most frequently used approach (25/36, 69.4%), followed by usability testing (16/36, 44.4%). End-user feedback sessions and surveys were reported in 22.2% (8/36) of studies, while focus groups appeared in 13.9% (5/36). Less commonly reported design methods included content analysis, scenarios, stakeholder consultations, and thematic analysis (2/36, 5.6% each). Participatory design sessions (2/36, 5.6%) and co-design workshops (1/36, 2.8%) were comparatively rare. Other approaches, such as direct observation, field trials, iterative design, journey mapping, pilot-testing, and system usage data, each appeared only once (1/36, 2.8% each).

Several cases illustrate this methodological skew toward consultative rather than collaborative techniques. *OncoKompa* [13] and *MijnAVL* [21] primarily drew on survivor interviews and usability testing to refine prototypes, while *ASyMS* [26] relied on structured usability evaluations, including think-aloud protocols, to improve device navigation. These approaches ensured tools were aligned with patient needs but often stopped short of deeper participatory design. By contrast, studies such as *StressProffen* [22] integrated stakeholder perspectives more systematically, combining evidence-based principles with design thinking methods to produce a culturally sensitive psychosocial intervention. Likewise, the *mPCL* [23] project in Tanzania demonstrated the value of participatory methods by actively involving patients, caregivers, and local health workers in shaping the platform, ensuring contextual relevance and cultural appropriateness.

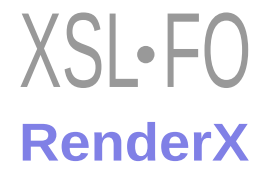

In summary, researcher-led design methods that capture user perspectives were predominant across the included studies, while collaborative and participatory approaches were less common. Only a small number of studies reported using co-design workshops, journey mapping, or multistakeholder sessions. This indicates that participatory techniques aimed at involving users and stakeholders throughout the design process remain limited in current cancer navigation research.

## Conceptualizations of Cancer Navigation Through UCD/HCD

### *Overview*

Across included studies, UCD/HCD approaches were most often applied as practical design strategies rather than overarching frameworks for navigation. Based on our findings from currently available literature, their use has been concentrated to three main functions: (1) evaluating usability and satisfaction outcomes as markers of navigation success, (2) centering patient perspectives as the primary driver of design decisions, and (3) using iterative cycles of prototyping and testing to refine tools. For example, O'Malley et al [29] developed and tested the *e-EXCELS* eHealth tool for cancer survivorship follow-up care.

Usability testing revealed concerns about patient burden, tool fatigue, timing of introduction, relevance, and privacy, underscoring the importance of using a UCD process.

Less frequently, studies broadened participation to include caregivers, clinicians, or navigators, or leveraged design methods to reconceptualize navigation itself as a process of information exchange, communication, and coordination.

In summary, UCD/HCD approaches in cancer navigation were mainly used to ensure that patient-facing tools are usable and acceptable. Fewer studies applied these approaches to address system-level outcomes, involve multiple stakeholders, or support implementation in real-world settings.

### *Outcomes as Navigation Constructs*

Across included studies, outcomes most often reflected usability (28/36, 77.8%) and user satisfaction (26/36, 72.2%), with these measures commonly serving as proxies for the overall success of cancer care navigational tools. Navigation effectiveness was reported less frequently (17/36, 47.2%), while patient engagement appeared in just over one-third of studies (13/36, 36.1%).

Broader outcomes such as clinical use (9/36, 25%), design or strategy insights (8/36, 22.2%), and health professional engagement (6/36, 16.7%) were less assessed. For example, Ho et al [23] evaluated a digital case management platform for community-based palliative care in sub-Saharan Africa by gathering clinicians' perspectives on its impact on care delivery, showing that it improved access, enabled remote symptom management, and supported more efficient patient prioritization. Even more rarely, studies addressed feasibility and implementation logistics (5/36, 13.9%), collaboration and coordination (3/36, 8.3%), or educational value (2/36, 5.6%). For example, Nikkhah et al [30] applied role theory to understand how parents act as caregiving teams in children's hospitals, emphasizing the need for design features that support coordination and shared responsibilities. As another example, Womack et al [31] engaged patients in conceptualizing clinic processes and digital health tools aimed at raising awareness and supporting education on integrative oncology services.

These findings reveal a strong emphasis on usability and user satisfaction in cancer navigation tools, illustrating how UCD/HCD has been applied primarily to ensure tools are acceptable and easy to use, rather than to measure their long-term effectiveness or system-level impact. For instance, *StressProffen* [22] and *Enable* [32] demonstrated high usability and satisfaction among survivors, underscoring the success of iterative, user-centered refinements. Similarly, *ASyMS* [26] usability testing revealed strong patient motivation to engage with symptom monitoring, even as participants called for a more attractive design and expanded features. Ahmadi et al [33] developed a mobile app for self-care management in Iranian women with breast cancer, and usability testing confirmed that its content and functions were appropriate and satisfactory. Moving beyond usability, tools such as *OncoKompas* [13] and *mPCL* [23] illustrate a shift toward evaluating navigation effectiveness and feasibility in real-world contexts, whether through tailoring survivorship resources or coordinating palliative care across diverse settings.

Overall, the reviewed studies primarily measured outcomes related to usability and user satisfaction, with fewer examining broader aspects such as navigation effectiveness, coordination, or sustainability. Only a limited number of studies assessed system-level or implementation-related outcomes, such as timeliness, collaboration, or equitable access. Toward this direction, Geerts et al [34] developed and evaluated *MM E-Coach*, a patient-centered eHealth application designed to support patients with multiple myeloma and health care professionals by improving treatment management, usability, and communication through an iterative, user- and stakeholder-driven design process. They further initiated a randomized clinical trial to evaluate its clinical effectiveness.

Similarly, Morse et al [35] designed, developed, and tested mPCL, a human-centered mobile and web application that supports outpatient palliative symptom assessment, care coordination, and pain control among Tanzanian patients with cancer by connecting patients, caregivers, local health workers, and palliative care specialists. Their findings emphasized the need for further evaluation to assess the tool's effectiveness and sustainability, particularly in remote and resource-limited settings. Mittal et al [27] also designed and deployed *SCOPE*, a platform for technology-enhanced collaborative care combining a patient-facing mobile app with a provider-facing registry. They deployed *SCOPE* for 24 months for actual care in 6 cancer clinics.

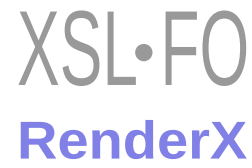

### Levels of Stakeholder Engagement in UCD/HCD

Across included studies, stakeholder engagement in UCD/HCD processes varied in scope and depth. Most studies engaged patients as the primary voice in the design process, often through semistructured interviews and surveys. This strong emphasis on patient perspectives reflects the value of empathy and coproduction, which are central principles of UCD/HCD as shown by Hou et al [36] and Womack et al [31]. Schadewaldt et al [37] also suggested that applying evidence and stakeholder input through a UCD/HCD approach can inform implementation strategies that reduce the risk of digital health failure.

In line with these findings, Glaser et al [38] evaluated the *CAPABLE* prototype for immunotherapy patients, showing that involving target users improved perceived usefulness and revealed critical symptom–reporting issues that would have gone unnoticed without their participation. Tools such as *OncoKompas* [13] and *MijnAVL* [21] exemplify this approach, with cancer survivors directly involved in iterative prototype testing to refine content, navigation features, and overall functionality. *CMyLife* presented by Ector et al [14], developed for individuals with chronic myeloid leukemia, equipped patients with tools to track and interpret their results while aligning design decisions with survivor priorities identified during participatory sessions. Similarly, *SmartSurvivor* by Baseman et al [39] explored mHealth options for survivorship care, explicitly embedding patient perspectives into the functionality and usability of the tool.

Although most included studies centered on participation of patients with cancer in UCD/HCD processes, a smaller but important subset extended participation to include health care providers, clinicians, caregivers, and navigators [40,41]. These examples demonstrate how human-centered approaches can extend beyond individual patients' needs to reflect the complex, multiactor systems that shape cancer care. For example, Salako et al [42] brought together researchers, clinicians, and technology developers to review remote symptom monitoring for palliative cancer care in Africa. They described 3 approaches from different countries, highlighting delivery models, outcomes, and implementation experiences. The study concluded that codeveloping and adapting these tools with local stakeholders is feasible, with user-centered insights guiding future scale-up.

Nabelsi and Plouffe [28] engaged nurse navigators and interfacility teams to refine a web-based coordination platform designed to improve oncology and nononcologic thoracic surgery care pathways. Similarly, the *mPCL* project by Ho et al [23] incorporated local health workers in Tanzania to co-design a mobile platform for palliative care, ensuring cultural and contextual relevance in a low-resource setting. Aronoff-Spencer et al [43] also described a human-centered participatory design approach that engaged patients, caregivers, and community stakeholders to support rural patients experiencing distress during cancer treatment. Patel et al [44] interviewed 20 caregivers of children with cancer to explore potential design solutions. They identified 4 caregiving coordination technology archetypes at diagnosis and showed how caregivers' preferences evolved over time.

Across the reviewed studies, inclusion of diverse stakeholders such as clinicians, caregivers, and navigators was less common compared to patient-only approaches. Studies that incorporated multiple stakeholder groups identified additional factors affecting navigation design, including workflow alignment, caregiver responsibilities, and infrastructure limitations. Overall, most studies maintained a primary focus on patient perspectives, with fewer addressing the broader system-level context of care delivery, such as organizational workflows, care coordination processes, technological infrastructure, and institutional constraints that shape how navigation services are implemented and sustained.

### Iteration and Responsiveness

Across included studies, UCD/HCD was frequently operationalized as an iterative cycle in which prototypes were tested and refined in response to user feedback. This responsiveness ensured that navigation tools were grounded in real-world needs and experiences.

For example, tools such as *StressProffen* [22] and *CMyLife* [14] demonstrated how usability testing can identify challenges with data entry and workflow guidance, which directly informed revisions to interface design and content presentation to improve navigation pathways. Likewise, iterative development of *MijnAVL* presented by Kuijpers et al [21] led to usability improvements, including the addition of FAQs and hyperlinks to improve patient experience. Similarly, an iterative development process of a cancer prevention web application by Carrera et al [45] and the prevention-focused web application described by Anders et al [32] incorporated repeated cycles of feedback that shaped tutorials and interactive features, supporting its potential to enhance risk understanding and healthy behavior.

### Redefining Navigation Through Design

Beyond creating tools for discrete functions, several studies applied UCD/HCD approaches to reconceptualize what "navigation" means in cancer care. Rather than limiting navigation to scheduling or wayfinding, studies framed it as a broader process of information exchange, communication, and care coordination.

For example, Basch et al [46] leveraged patient-reported outcomes to enhance communication between patients and clinicians, transforming symptom monitoring into a navigation mechanism for timely intervention and treatment adjustment. Similarly, Donawa et al [47] co-designed a mobile tool for distress management in rural communities, positioning navigation as an integrated support system that addressed emotional, informational, and logistical barriers simultaneously. *Interaktor,* presented by Langius-Eklöf et al [48], also broadened navigation by increasing patients' sense of security and self-reflection, thereby embedding psychosocial support directly into routine symptom monitoring.

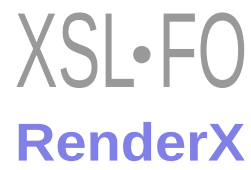

# Discussion

## Principal Findings

This scoping review synthesizes how UCD and HCD principles have been applied in the development and implementation of digital cancer navigation, while also identifying gaps that limit broader impact. Overall, UCD/HCD was most often applied to patient-facing tools during prototyping and usability evaluation, with comparatively less attention to early discovery phases and downstream implementation in routine care settings. These efforts appear to have improved alignment between navigational tools and patient needs; however, fewer studies addressed system-level barriers, equity, cancer-specific navigational needs, and global applicability. Future research and development should consider deliberately extending UCD/HCD across a wider variety of cancers, diverse geographic contexts, multistakeholder design processes, and outcomes that measure more than usability. To support interpretive clarity, we distinguish between findings derived directly from the reviewed studies and forward-looking implications intended to inform future research, design practice, and implementation.

Findings demonstrate that UCD/HCD in cancer navigation has been applied most extensively to adult cancers in high-income settings. Pediatric and rare cancers were comparatively underrepresented, as were studies conducted in low- and middle-income settings, despite the substantial coordination demands often present in these contexts. Addressing these gaps may be important for improving the generalizability and equity relevance of digital navigation research. As an illustrative scenario, in a pediatric oncology ward in sub-Saharan Africa, co-design workshops with caregivers could reveal the daily burdens of medication access and transportation [23]. Embedding those insights into a navigation app might prioritize transportation support (eg, ride-share scheduling or community volunteer coordination) alongside education, addressing barriers that may influence treatment continuity. Without attention to these barriers, digital navigation may risk reproducing existing inequities rather than mitigating them.

At the stakeholder level, most tools were developed directly with patient input, demonstrating the strength of UCD/HCD in capturing and centering lived experience. However, caregivers, navigators, and clinicians were rarely included, which may leave workflow and coordination barriers underexplored. Caregivers and patient navigators can play pivotal roles in access, treatment timeliness, and continuity of cancer care, while clinician engagement at each iterative stage of UCD/HCD is critical for integration into workflows and sustained adoption [2,30]. Limiting participatory design primarily to patients could risk overlooking system-level and workforce perspectives needed for real-world implementation. To address this gap, design processes could prioritize multistakeholder participation. For instance (as an illustrative example), in developing a survivorship platform for rural patients, engaging nurse navigators could identify how symptom alerts fit within clinical workflows, while caregiver voices could flag gaps in home-based support. This approach may shift tools from stand-alone patient tools toward integrated systems designed to bolster continuity of care.

Current tools focus heavily on education and information provision, with limited attention to psychosocial, financial, or logistical barriers. These functions may improve self-efficacy and patient activation [49], but they may be insufficient for addressing the psychosocial, financial, or logistical barriers that many patients face [50]. While UCD/HCD approaches have supported the development of education and communication functions responsive to patient needs, future applications could also extend to the broader spectrum of navigation services that shape equity, access, and continuity of care. This pattern reflects how UCD/HCD has been applied primarily to informational and communicative aspects of navigation, while more complex or resource-intensive functions remain underexplored.

## Strengths and Limitations

This review offers one of the first comprehensive maps of how UCD/HCD principles have been applied to digital and mHealth tools for navigation for patients with cancer. A comprehensive search strategy was applied across 7 major databases, combining controlled vocabulary and free-text terms to capture diverse design approaches and navigation functions. However, several limitations should be noted. Despite the breadth of databases searched, the review was limited to English-language publications between 2015 and 2025, potentially excluding relevant studies in other languages or earlier foundational work. While the dual-review process enhanced rigor, inclusion criteria required explicit mention of UCD or HCD, which may have excluded studies using participatory approaches under different terminology. Although descriptive and thematic analysis revealed important patterns, no formal quality appraisal was conducted, in keeping with scoping review methodology. Given this, the strength of evidence underlying individual findings could not be assessed. Moreover, most studies were conducted in high-income countries and focused on adult cancer populations, limiting generalizability to low- and middle-income settings or underrepresented cancer types.

## Future Directions

Overall, the review’s comprehensive scope provides a foundation for future research. The implications outlined below are forward-looking and intended to inform future research and practice; several examples are illustrative and were not directly evaluated in the reviewed studies. Subsequent systematic reviews should incorporate quality appraisal frameworks, cross-cultural analyses, and longitudinal evaluations to further examine the effectiveness, equity impact, and sustainability of UCD/HCD in cancer navigation.

Moving forward, developers should also consider expanding navigation functions to directly address psychosocial, financial, and logistical barriers identified in the literature. For example, a survivorship tool could embed a financial navigation module where patients track insurance claims and receive cost-of-living support resources, similar to the Wheeler et al [51] model. Likewise, a peritreatment app might integrate transportation (ie, ride-share) scheduling to reduce missed appointments in gynecologic oncology. Designing and implementing these functions may require moving beyond informational features toward more complex domains with potential system-level impact.

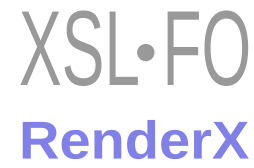

Most reviewed studies relied on traditional UCD/HCD design methods (ie, interviews, surveys, and usability testing applied midcycle) with limited reporting of upstream discovery or downstream implementation. Discovery and routine care implementation were rarely discussed, which may limit both innovation and sustainability. This pattern reflects broader digital health literature, where many interventions are tested in laboratory- or pilot-type settings, very few report robust real-world deployment, maintenance, or adoption over time [52]. Overreliance on consultative design methods risks capturing only surface-level usability preferences while overlooking systemic challenges such as workflow misalignments, caregiver burdens, or sociotechnical barriers [53]. These observations and gaps carry important public health policy implications, as insufficient integration of UCD/HCD across the design-to-implementation continuum can undermine investments in digital oncology infrastructure, limit scalability, and hinder equitable access to effective navigation technologies. Health systems and policymakers should prioritize frameworks that embed UCD/HCD into digital health standards and implementation incentives.

Additionally, a shift toward embedding participatory and iterative design methods from start to finish is needed. Journey mapping at the discovery stage could uncover bottlenecks in care transitions (ie, from surgery to rehabilitation), guiding more effective tool design. Co-design workshops and multistakeholder field trials during implementation could then test how tools integrate into routine care, adapt across cancer contexts, and endure beyond pilots into real-world applications [54-57].

Correspondingly, outcomes were dominated by usability and satisfaction, while far fewer studies assessed navigation effectiveness, care coordination, or clinical use. Moreover, even fewer included measures of equity, timeliness, or continuity of care. This narrow scope risks producing tools that are usable but not transformative for access, coordination, or sustainability. Evaluation frameworks must expand to capture system-level outcomes. Pairing implementation outcomes with theory-informed, causally grounded evaluation approaches can help assess not only acceptability, but whether proposed mechanisms (eg, improved coordination leading to fewer delays) are supported by observed data rather than usability proxies alone [58].

One example could include comprehensive trials to assess whether navigation apps reduce treatment delays across racial groups, improve communication between community hospitals and tertiary centers, or decrease financial burden for patients and families. Embedding these outcomes would help ensure UCD/HCD advances are judged by their tangible contribution to health equity and system performance, and not just whether users like the interface.

Finally, UCD/HCD approaches may contribute to broadening the conceptualization of navigation from wayfinding toward communication, coordination, and empowerment. Tools such as *StressProffen* [22] and *CMyLife* [14] illustrate how iterative design can expand navigation beyond tracking to include psychosocial support and relational care. For practice, this shift underscores the importance of embedding patient perspectives alongside those of clinicians, navigators, and caregivers, ensuring tools align with real-world workflows. Scaling this orientation across cancer contexts and diverse health systems could transform navigation from a patient app to a core function of oncology care delivery. For research, future studies should move beyond single-site, short-term evaluations toward longitudinal and multicountry designs that capture not only usability but also system-level outcomes. By expanding participatory approaches and testing tools in diverse contexts, UCD/HCD has the potential to transform navigation into a durable, equity-sensitive component of cancer care delivery.

## Conclusions

UCD and HCD approaches have demonstrated value in shaping digital cancer navigation tools that are usable, acceptable, and responsive to patient needs; however, their current application remains narrow, concentrated in adult cancers, high-income settings, and informational functions. While these contributions are notable, relatively few studies address the broader coordination, equity, and implementation challenges that define navigation in real-world oncology settings.

The evidence from this review suggests that UCD/HCD has been most impactful when applied iteratively, embedding patient perspectives to refine tools that improve engagement, communication, and self-management. At the same time, caregiver, navigator, and clinician voices remain underrepresented, participatory methods are underused, and outcomes largely privilege usability over system-level effectiveness. These patterns may limit both the scalability and sustainability of current tools. To more fully realize the full potential of UCD/HCD in cancer navigation, future research and practice should extend beyond interface refinement toward the development of equitable, context-sensitive systems. This includes deliberate inclusion of a wide variety of cancers and low- and middle-income settings, broader stakeholder engagement across patients, caregivers, and providers, and systematic use of participatory and co-design methods that span discovery through implementation.

Equally, evaluation frameworks should extend beyond usability to capture coordination, timeliness, continuity, and equity as essential markers of navigation success. These findings underscore that UCD/HCD should both refine cancer navigation tools and redefine what navigation means within oncology. An integrated process of empowerment, communication, and coordination across diverse care contexts is warranted. By embracing inclusivity, methodological innovation, and system-level evaluation, the next generation of digital navigational tools can move beyond promising prototypes to sustainable solutions that improve outcomes and reduce disparities in cancer care delivery.

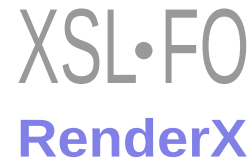

## Funding

This study was supported by a grant from the State of Tennessee Department of Health.

## Data Availability

Data sharing is not applicable to this article as no new data were created or analyzed in this study. All data extracted and analyzed in this review are available in Multimedia Appendix 2, which provides bibliographic and methodological details of the included studies.

## Authors' Contributions

Conceptualization: SK, BMW
Drafting of manuscript: SK, BMW, AS-N
Review and editing of manuscript: SK, BMW, PKR, JAZ, SH, FAK, RR, CLB, RLD, DLS, AS-N
Supervision: DLS, AS-N
Funding acquisition: DLS, AS-N

## Conflicts of Interest

None declared.

## Multimedia Appendix 1

PRISMA-ScR checklist.
[DOCX File , 84 KB-Multimedia Appendix 1]

## Multimedia Appendix 2

Summary of Included Studies and Full Search Strategies.
[DOCX File , 582 KB-Multimedia Appendix 2]

## References

1. Bray F, Laversanne M, Sung H, Ferlay J, Siegel RL, Soerjomataram I, et al. Global cancer statistics 2022: GLOBOCAN estimates of incidence and mortality worldwide for 36 cancers in 185 countries. CA Cancer J Clin. 2024;74(3):229-263. [FREE Full text] [doi: 10.3322/caac.21834] [Medline: 38572751]
2. Chen M, Wu VS, Falk D, Cheatham C, Cullen J, Hoehn R. Patient navigation in cancer treatment: a systematic review. Curr Oncol Rep. 2024;26(5):504-537. [FREE Full text] [doi: 10.1007/s11912-024-01514-9] [Medline: 38581470]
3. Freeman HP, Rodriguez RL. History and principles of patient navigation. Cancer. 2011;117(15 Suppl):3539-3542. [FREE Full text] [doi: 10.1002/cncr.26262] [Medline: 21780088]
4. Sleiman M, Yockel M, Fleischmann A, Silber E, Liu M, Young O, et al. The role of peer support and patient navigation for empowerment in breast cancer survivors: implications for community cancer control. J Psychosoc Oncol Res Pract. 2024;6(2):128. [doi: 10.1097/or9.0000000000000128] [Medline: 39463701]
5. Neal CD, Weaver DT, Raphel TJ, Lietz AP, Flores EJ, Percac-Lima S, et al. Patient navigation to improve cancer screening in underserved populations: reported experiences, opportunities, and challenges. J Am Coll Radiol. 2018;15(11):1565-1572. [doi: 10.1016/j.jacr.2018.03.001] [Medline: 29685346]
6. Maita KC, Maniaci MJ, Haider CR, Avila FR, Torres-Guzman RA, Borna S, et al. The impact of digital health solutions on bridging the health care gap in rural areas: a scoping review. Perm J. 2024;28(3):130-143. [FREE Full text] [doi: 10.7812/TPP/23.134] [Medline: 39135461]
7. Monjaras-Avila C, Blakaj D, Hu K, Yamoah K, Gamez E, Comas A, et al. Integrating digital health into oncology: a comprehensive review. Preprints.org. Preprint posted online on January 7, 2026. 2026. [doi: 10.20944/preprints202601.0502.v1]
8. Kheirinejad S, Visuri A, Suryanarayana SA, Hosio S. Exploring mHealth applications for self-management of chronic low back pain: a survey of features and benefits. Heliyon. 2023;9(6):e16586. [FREE Full text] [doi: 10.1016/j.heliyon.2023.e16586] [Medline: 37346357]
9. Ngo V, Matsumoto CG, Joseph JG, Bell JF, Bold RJ, Davis A, et al. The personal health network mobile app for chemotherapy care coordination: qualitative evaluation of a randomized clinical trial. JMIR Mhealth Uhealth. 2020;8(5):e16527. [FREE Full text] [doi: 10.2196/16527] [Medline: 32452814]
10. Anderson NS, Norman DA, Draper SW. User Centered System Design: New Perspectives on Human-Computer Interaction. Hillsdale, NJ. Lawrence Erlbaum Associates; 1986.
11. Melles M, Albayrak A, Goossens R. Innovating health care: key characteristics of human-centered design. Int J Qual Health Care. 2021;33(Supplement_1):37-44. [FREE Full text] [doi: 10.1093/intqhc/mzaa127] [Medline: 33068104]

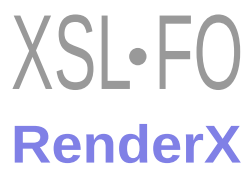

12. Carr AL, Vinod N, Farha P, Lu T, Farha MJ, Graves KD. An exploration of facilitators and barriers to patient navigator core functions with breast cancer patients: implications for the development of a human-centered mHealth app. PEC Innov. 2023;3:100226. [FREE Full text] [doi: 10.1016/j.pecinn.2023.100226] [Medline: 37859864]
13. Duman-Lubberding S, van Uden-Kraan CF, Jansen F, Witte BI, van der Velden LA, Lacko M, et al. Feasibility of an eHealth application "OncoKompas" to improve personalized survivorship cancer care. Support Care Cancer. 2016;24(5):2163-2171. [FREE Full text] [doi: 10.1007/s00520-015-3004-2] [Medline: 26563178]
14. Ector GI, Westerweel PE, Hermens RP, Braspenning KA, Heeren BC, Vinck OM, et al. The development of a web-based, patient-centered intervention for patients with chronic myeloid leukemia (CMyLife): design thinking development approach. J Med Internet Res. 2020;22(5):e15895. [FREE Full text] [doi: 10.2196/15895] [Medline: 32412424]
15. Santos WMD, Secoli SR, Püschel V. The Joanna Briggs Institute approach for systematic reviews. Rev Lat Am Enfermagem. 2018;26:e3074. [FREE Full text] [doi: 10.1590/1518-8345.2885.3074] [Medline: 30462787]
16. Tricco AC, Lillie E, Zarin W, O'Brien KK, Colquhoun H, Levac D, et al. PRISMA extension for scoping reviews (PRISMA-ScR): checklist and explanation. Ann Intern Med. 2018;169(7):467-473. [FREE Full text] [doi: 10.7326/M18-0850] [Medline: 30178033]
17. Covidence systematic review software. Veritas Health Innovation. URL: https://www.covidence.org/ [accessed 2026-04-08]
18. Mendu S, Boukhechba M, Gordon J, Datta D, Molina E, Arroyo G, et al. Design of a culturally-informed virtual human for educating Hispanic women about cervical cancer. Int Conf Pervasive Comput Technol Healthc. 2018;2018:360-366. [FREE Full text] [doi: 10.1145/3240925.3240968] [Medline: 30555731]
19. Kim K, Bell J, Bold R. A personal health network for chemotherapy care coordinationvaluation of usability among patients. Stud Health Technol Inform. 2016;225:232-236. [doi: 10.3233/978-1-61499-658-3-232]
20. Mendes-Santos C, Nunes F, Weiderpass E, Santana R, Andersson G. Development and evaluation of the usefulness, usability, and feasibility of iNNOV breast cancer: mixed methods study. JMIR Cancer. 2022;8(1):e33550. [FREE Full text] [doi: 10.2196/33550] [Medline: 35166682]
21. Kuijpers W, Groen WG, Oldenburg HS, Wouters MW, Aaronson NK, van Harten WH. Development of MijnAVL, an interactive portal to empower breast and lung cancer survivors: an iterative, multi-stakeholder approach. JMIR Res Protoc. 2015;4(1):e14. [FREE Full text] [doi: 10.2196/resprot.3796] [Medline: 25614924]
22. Børøsund E, Mirkovic J, Clark MM, Ehlers SL, Andrykowski MA, Bergland A, et al. A stress management app intervention for cancer survivors: design, development, and usability testing. JMIR Form Res. 2018;2(2):e19. [FREE Full text] [doi: 10.2196/formative.9954] [Medline: 30684438]
23. Ho YX, Morse RS, Lambden K, Mushi BP, Ngoma M, Mahuna H, et al. How a digital case management platform affects community-based palliative care of Sub-Saharan African cancer patients: clinician-users' perspectives. Appl Clin Inform. 2022;13(5):1092-1099. [FREE Full text] [doi: 10.1055/s-0042-1758223] [Medline: 36384234]
24. Ankrah E, Bhattacharya A, Donjuan L. When worlds collide: boundary management of adolescent and young adult childhood cancer survivors and caregivers. 2022. Presented at: CHI '22: Proceedings of the 2022 CHI Conference on Human Factors in Computing Systems; April 29:1-16; New Orleans, LA, USA. [doi: 10.1145/3491102.3517544]
25. Lau GJ, Loiselle CG. E-health tools in oncology nursing: perceptions of nurses and contributions to patient care and advanced practice. Can Oncol Nurs J. 2018;28(2):118-124. [FREE Full text] [doi: 10.5737/23688076282118124] [Medline: 31148816]
26. Moradian S, Krzyzanowska MK, Maguire R, Morita PP, Kukreti V, Avery J, et al. Usability evaluation of a mobile phone-based system for remote monitoring and management of chemotherapy-related side effects in cancer patients: mixed-methods study. JMIR Cancer. 2018;4(2):e10932. [FREE Full text] [doi: 10.2196/10932] [Medline: 30578238]
27. Mittal A, Jones T, Karkar R, Suh J, Williams S, Zheng Y, et al. SCOPE: examining technology-enhanced collaborative care management of depression in the cancer setting. Proc ACM Hum Comput Interact. 2025;9(2):CSCW162. [doi: 10.1145/3711060] [Medline: 40519326]
28. Nabelsi V, Plouffe V. Enhancing care coordination in oncology and nononcology thoracic surgery care pathways through a digital health solution: mixed methods study. JMIR Form Res. 2024;8:e60222. [FREE Full text] [doi: 10.2196/60222] [Medline: 39591606]
29. O'Malley DM, Davis SN, Devine KA, Sullivan B, Bator A, Clemow L, et al. Development and usability testing of the e-EXCELS tool to guide cancer survivorship follow-up care. Psychooncology. 2020;29(1):123-131. [FREE Full text] [doi: 10.1002/pon.5222] [Medline: 31626397]
30. Nikkhah S, Rode AU, Mittal P. "I feel like I need to split myself in half": using role theory to design for parents as caregiving teams in the children's hospital. 2022. Presented at: CSCW'22 Companion: Companion Computer Supported Cooperative Work and Social Computing; November 8-22:115-120; Virtual Event, Taiwan. [doi: 10.1145/3500868.3559466]
31. Womack DM, Kennedy R, Chamberlin SR, Rademacher AL, Sliney CD. Patients' lived experiences and recommendations for enhanced awareness and use of integrative oncology services in cancer care. Patient Educ Couns. 2022;105(7):2557-2561. [FREE Full text] [doi: 10.1016/j.pec.2021.11.018] [Medline: 34865887]
32. Anders C, Moorthy P, Svensson L, Müller J, Heinze O, Knaup P, et al. Usability and user experience of an mHealth app for therapy support of patients with breast cancer: mixed methods study using eye tracking. JMIR Hum Factors. 2024;11:e50926. [FREE Full text] [doi: 10.2196/50926] [Medline: 38441959]

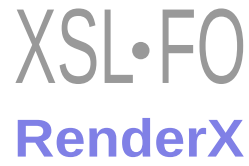

33. Ahmadi M, Shahrokhi SN, Khavaninzadeh M, Alipour J. Development of a mobile-based self-care application for patients with breast cancer-related lymphedema in Iran. Appl Clin Inform. 2022;13(5):935-948. [FREE Full text] [doi: 10.1055/s-0042-1757295] [Medline: 36198310]
34. Geerts P, Eijsink J, Moser A, Ter Horst P, Boersma C, Postma M. Rationale and development of an e-health application to deliver patient-centered care during treatment for recently diagnosed multiple myeloma patients: pilot study of the MM E-coach. Pilot Feasibility Stud. 2023;9(1):85. [FREE Full text] [doi: 10.1186/s40814-023-01307-0] [Medline: 37210584]
35. Morse RS, Lambden K, Quinn E, Ngoma T, Mushi B, Ho YX, et al. A mobile app to improve symptom control and information exchange among specialists and local health workers treating Tanzanian cancer patients: human-centered design approach. JMIR Cancer. 2021;7(1):e24062. [FREE Full text] [doi: 10.2196/24062] [Medline: 33755022]
36. Hou IC, Lan MF, Shen SH, Tsai PY, Chang KJ, Tai HC, et al. The development of a mobile health app for breast cancer self-management support in Taiwan: design thinking approach. JMIR Mhealth Uhealth. 2020;8(4):e15780. [FREE Full text] [doi: 10.2196/15780] [Medline: 32352390]
37. Schadewaldt V, O'Brien T, Kalla M, Krishnasamy M, Burns K, Bray SC, et al. Development of an evidence-informed implementation strategy for a digital supportive care platform for brain tumour patients, their carers and healthcare professionals. Digit Health. 2025;11:20552076251316713. [FREE Full text] [doi: 10.1177/20552076251316713] [Medline: 39949847]
38. Glaser SLC, Fraterman I, van Brummelen N, Tibollo V, Del Campo LM, Mallo H, et al. Usability and usefulness of a symptom management coaching system for patients with cancer treated with immune checkpoint inhibitors: comparative mixed methods study. JMIR Form Res. 2025;9:e57659. [FREE Full text] [doi: 10.2196/57659] [Medline: 39847771]
39. Baseman J, Revere D, Baldwin LM. A mobile breast cancer survivorship care app: pilot study. JMIR Cancer. 2017;3(2):e14. [FREE Full text] [doi: 10.2196/cancer.8192] [Medline: 28951383]
40. Timmerman JG, Tönis TM, Dekker-van Weering MGH, Stuiver MM, Wouters MWJM, van Harten WH, et al. Co-creation of an ICT-supported cancer rehabilitation application for resected lung cancer survivors: design and evaluation. BMC Health Serv Res. 2016;16:155. [FREE Full text] [doi: 10.1186/s12913-016-1385-7] [Medline: 27121869]
41. Iott B, Caverly T, Fishstrom A, King D, Meng G, Flynn A. Clinician perspectives on the user experience, configuration, and scope of use of a patient reported outcomes (PRO) dashboard. 2019. Presented at: PervasiveHealth'19: Proceedings of the 13th EAI International Conference on Pervasive Computing Technologies for Healthcare; May 20-23:21-30; Trento Italy. [doi: 10.1145/3329189.3329198]
42. Salako O, Enyi A, Miesfeldt S, Kabukye JK, Ngoma M, Namisango E, et al. Remote symptom monitoring to enhance the delivery of palliative cancer care in low-resource settings: emerging approaches from Africa. Int J Environ Res Public Health. 2023;20(24):7190. [FREE Full text] [doi: 10.3390/ijerph20247190] [Medline: 38131741]
43. Aronoff-Spencer E, McComsey MY, Chih M, Hubenko A, Baker C, Kim J, et al. Designing a framework for remote cancer care through community co-design: participatory development study. J Med Internet Res. 2022;24(4):e29492. [FREE Full text] [doi: 10.2196/29492] [Medline: 35412457]
44. Patel S, Nikkhah S, Nadkarni S. Designing for resilient caregiving coordination journeys: prioritizing features across families and phases. 2024. Presented at: CSCW Companion '24: Companion Publication of the 2024 Conference on Computer-Supported Cooperative Work and Social Computing; 2024 November 9-13:525-532; San Jose Costa Rica. [doi: 10.1145/3678884.3681901]
45. Carrera P, Elias O, Zhang R. User-centered design and usability evaluation of a cancer prevention web application: an iterative approach in Germany. medRxiv. Preprint posted online on April 30, 2025. 2025. [doi: 10.1101/2025.04.29.25326679]
46. Basch E, Stover AM, Schrag D, Chung A, Jansen J, Henson S, et al. Clinical utility and user perceptions of a digital system for electronic patient-reported symptom monitoring during routine cancer care: findings from the PRO-TECT trial. JCO Clin Cancer Inform. 2020;4:947-957. [FREE Full text] [doi: 10.1200/CCI.20.00081] [Medline: 33112661]
47. Donawa A, Powell C, Wang R, Chih MY, Patel R, Zinner R, et al. Designing survey-based mobile interfaces for rural patients with cancer using Apple's ResearchKit and CareKit: usability study. JMIR Form Res. 2024;8:e57801. [FREE Full text] [doi: 10.2196/57801] [Medline: 39326043]
48. Langius-Eklöf A, Christiansen M, Lindström V, Blomberg K, Hälleberg Nyman M, Wengström Y, et al. Adherence to report and patient perception of an interactive app for managing symptoms during radiotherapy for prostate cancer: descriptive study of logged and interview data. JMIR Cancer. 2017;3(2):e18. [FREE Full text] [doi: 10.2196/cancer.7599] [Medline: 29089290]
49. Tuominen L, Leino-Kilpi H, Poraharju J, Cabutto D, Carrion C, Lehtiö L, et al. Interactive digital tools to support empowerment of people with cancer: a systematic literature review. Support Care Cancer. 2024;32(6):396. [FREE Full text] [doi: 10.1007/s00520-024-08545-9] [Medline: 38816629]
50. Dau H, Kassam F, Payne BA, Miller H, Ogilvie G. Digital technology as a tool to provide social support to individuals with cancer in low- and middle-income countries: a scoping review. PLOS Digit Health. 2025;4(4):e0000609. [doi: 10.1371/journal.pdig.0000609] [Medline: 40257979]
51. Wheeler SB, Manning ML, Gellin M, Padilla N, Spees LP, Biddell CB, et al. Impact of a comprehensive financial navigation intervention to reduce cancer-related financial toxicity. J Natl Compr Canc Netw. 2024;22(8):557-562. [doi: 10.6004/jnccn.2024.7030] [Medline: 39413831]

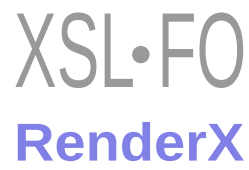

52. Duffy A, Christie GJ, Moreno S. The challenges toward real-world implementation of digital health design approaches: narrative review. JMIR Hum Factors. 2022;9(3):e35693. [FREE Full text] [doi: 10.2196/35693] [Medline: 36083628]
53. Busolo D, Woodgate R. Palliative care experiences of adult cancer patients from ethnocultural groups: a qualitative systematic review protocol. JBI Database System Rev Implement Rep. 2015;13(1):99-111. [doi: 10.11124/jbisrir-2015-1809] [Medline: 26447011]
54. Voorheis P, Wong JV, Lazarevic N, Imtiaz B, Bhuiya A, Steele Gray C. Using journey mapping and service blueprinting to design digital health behavior change innovations: a scoping review. J Diabetes Sci Technol. 2025;19(5):1196-1214. [doi: 10.1177/19322968251334396] [Medline: 40338057]
55. Messina A, Annoni AM, Amati R, Bano B, Franscella G, Albanese E, et al. Participatory methods in designing digital health interventions for informal caregivers of people with dementia. A systematic review. Internet Interv. 2025;39:100799. [FREE Full text] [doi: 10.1016/j.invent.2024.100799] [Medline: 40161473]
56. Henni SH, Maurud S, Fuglerud KS, Moen A. The experiences, needs and barriers of people with impairments related to usability and accessibility of digital health solutions, levels of involvement in the design process and strategies for participatory and universal design: a scoping review. BMC Public Health. 2022;22(1):35. [FREE Full text] [doi: 10.1186/s12889-021-12393-1] [Medline: 34991540]
57. Jallow F, Orbegozo AM, Caruhapoma MDC. Design workshops: participatory co-design and iterative adaptation of cervical cancer screening implementation plans in Peru. Cancer Epidemiol Biomarkers Prev. 2023;32((6_Suppl)):28. [doi: 10.1158/1538-7755.asgcr23-abstract-28]
58. Brenas JH, Shaban-Nejad A. Health intervention evaluation using semantic explainability and causal reasoning. IEEE Access. 2020;8:9942-9952. [doi: 10.1109/access.2020.2964802]

## Abbreviations

**HCD:** human-centered design
**mHealth:** mobile health
**PRISMA-ScR:** Preferred Reporting Items for Systematic Reviews and Meta-Analyses extension for Scoping Reviews
**UCD:** user-centered design

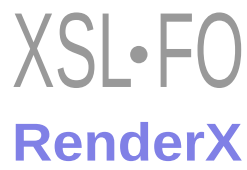